\documentclass[11pt]{amsart}

\usepackage{amsmath,amsfonts,amssymb}
\usepackage[dvips]{graphicx}
\usepackage{color}
\usepackage[square,sort&compress]{natbib}

\graphicspath{{./}{./figs/}}

\newtheorem{theorem}{Theorem}[section]   
\theoremstyle{definition}
\newtheorem{definition}[theorem]{Definition}   
\theoremstyle{remark}
\newtheorem{example}[theorem]{Example}        
\numberwithin{equation}{section}     

\long\def\Comment#1{
{
 \endIgnore{\relax} 
 }}
\def\endIgnore{\relax}

\begin{document}

\title
{Design of experiments and biochemical network inference }

\author[R.~Laubenbacher]{Reinhard~Laubenbacher}
\address{Virginia Bioinformatics Institute\\
Virginia Polytechnic Institute and State University} 
\email[Reinhard~Laubenbacher]{reinhard@vbi.vt.edu}

\author[B.~Stigler]{Brandilyn~Stigler}
\address{Mathematical Biosciences Institute\\%
The Ohio State University}
\email[Brandilyn~Stigler]{bstigler@mbi.osu.edu}


\date{\today}


\begin{abstract}
Design of experiments is a branch of statistics that aims to
identify efficient procedures for planning experiments in order to
optimize knowledge discovery.  Network inference is a subfield of
systems biology devoted to the identification of biochemical
networks from experimental data. Common to both areas of research
is their focus on the maximization of information gathered from
experimentation. The goal of this paper is to establish a
connection between these two areas coming from the common use of
polynomial models and techniques from computational algebra.

\end{abstract}

\maketitle


\section{Introduction}
Originally introduced in \cite{PRW}, the field of \emph{algebraic
statistics} focuses on the application of techniques from
computational algebra and algebraic geometry to problems in
statistics.  One initial focus of the field was the design of
experiments, beginning with \cite{PW, Ri}. An early exposition
of a basic mathematical relationship between problems in the
design of experiments and computational commutative algebra
appeared in \cite{Ro}.  The basic strategy of \cite{Ro} and other
works is to
construct an algebraic model, in the form of a polynomial function
with rational coefficients, of a fractional factorial design. The
variables of the polynomial function correspond to the factors of
the design.  One can then use algorithmic techniques from
computational commutative algebra to answer a variety of
questions, for instance about the classification of all polynomial
models that are identified by a fractional design.

If $\mathbf p_1, \ldots , \mathbf p_r$ are the points of a
fractional design with $n$ levels, then the key algebraic object
to be considered is the \emph{ideal of points} $I$ that contains
all polynomials with rational coefficients that vanish on all
$\mathbf p_i$.  (See the appendix for a review of basic concepts
from commutative algebra.) The form of the polynomials in
different generating sets of this ideal is of special interest.
In particular, we are interested in so-called interpolator polynomials
which have a unique representation, given an explicit choice of generating set.
An interpolator polynomial $f(x_1,\ldots ,x_n)$ has the property that if $b_1,\ldots ,b_r$
is a response to the design given by the $\mathbf p_i$, then
$f(\mathbf p_i)=b_i$.

Strikingly similar constructions have been used recently to solve an
entirely different set of problems related to the inference of
intracellular biochemical networks, such as gene regulatory
networks, from experimental observations.  Relatively recent
technological breakthroughs in molecular biology have made possible
the simultaneous measurement of many different biochemical species
in cell extracts.  For instance, using DNA microarrays one can
measure the concentration of mRNA molecules, which provide
information about the activity levels of the corresponding genes at
the time the cell extract was prepared. Such network-level
measurements provide the opportunity to construct large-scale models
of molecular systems, including gene regulatory networks.

Here, an experimental observation consists of the measurement of
$n$ different quantities at $r$ successive time points, resulting
in a time course of $n$-dimensional real-valued vectors $\mathbf
p_1, \ldots ,\mathbf p_r$.  The number $r$ of experimental
observations is typically very small compared to the number $n$ of
quantities measured, due in part to the considerable expense of
making measurements. In recent years there has been tremendous
research activity devoted to the development of mathematical and
statistical tools to infer the entire network structure from such
a limited set of experimental measurements.

Inferring networks from data is a central problem in computational
systems biology, and several approaches have been developed using
a variety of approaches.  Models range from statistical models
such as Bayesian networks to dynamic models such as Markov chains
and systems of differential equations.  Another modeling framework
is that of finite dynamical systems such as Boolean networks.  A
method proposed in \cite{LS} uses such data to construct a
multi-state discrete dynamical system
$$ f = (f_1,\ldots ,f_n):k^n\longrightarrow k^n
$$
over a
finite field $k$
such that the coordinate functions $f_i$ are polynomials in
variables $x_1,\ldots ,x_n$ corresponding to the $n$ biochemical
compounds measured.  The system $f$ has to fit the given time course
data set, that is, $f(\mathbf p_i)=\mathbf p_{i+1}$ for $i=1,\ldots
, r-1$.  The goal is to infer a ``best'' or most likely model~$f$
from a given data set which specifies a fraction of the possible
state transitions of~$f$.  An advantage to working in a finite field is
that all functions $k^n \rightarrow k$ are represented by polynomials.
An important, and unanswered, question is
to design biological experiments in an optimal way in order to infer
a likely model with high probability.  One complicating factor is
that biochemical networks tend to be highly nonlinear.

In this paper, we describe the two approaches and point out the
similarities between the two classes of problems, the techniques
used to solve them, and the types of questions asked.
\section{Design of experiments}
In this section we provide a description of the computational
algebra approach to experimental design given in \cite{Ro, PRW}.
Let $\mathcal D$ be the full factorial design with $n$ factors. We
make the additional simplifying assumptions that each factor has
the same number $p$ of levels, resulting in $p^n$ points for~$\mathcal D$.
A {\it model} for the design is a function
$$
f:\mathcal D\longrightarrow \mathbb Q,
$$
that is, $f$ maps each point of~$\mathcal D$ to a measurement.
Instead of using the field $\mathbb Q$ for measurements, one may
choose other fields such as $\mathbb C$ or a finite field.  From
here on we will denote the field by $k$. It is well-known that any
function from a finite number of points in $k^n$ to $ k$ can be
represented by a polynomial, so we may assume that $f$ is a
polynomial in variables $x_1,\ldots , x_n$ with coefficients in $k$.

\begin{definition} A subset $\mathcal F =\{\mathbf p_1, \ldots ,
\mathbf p_r\}\subset \mathcal D$ is called a \emph{fraction}
of~$\mathcal D$.
\end{definition}

We list three important problems in the design of
experiments:
\begin{enumerate}
    \item Identify a model for the full design $\mathcal D$
    from a suitably chosen fraction~$\mathcal F$.
    \item Given information about features of the model, such as a list
    of the \emph{monomials} (power products) appearing in it, design a fraction~$\mathcal
    F$ which identifies a model for $\mathcal D$ with these features.
    \item Given a fraction $\mathcal F$, which models can
    be identified by it?
\end{enumerate}

These problems can be formulated in the language of computational
algebra making them amenable to solution by techniques from this
field.  The fraction $\mathcal F$ is encoded by an algebraic
object $I(\mathcal F)$, an ideal in the polynomial ring $k[x_1,
\ldots ,x_n]$.  This ideal contains all those polynomial functions
$g\in k[x_1, \ldots ,x_n]$ such that $g(\mathbf p_i)=0$ for all
$i=1, \ldots , r$.  It is called the \emph{ideal of points} of the
$\mathbf p_i$ and contains all polynomials \emph{confounded by}
the points in~$\mathcal F$.  Here we assume that the points are
distinct. We will see that one can draw
conclusions about $\mathcal F$ from its ideal of confounding
polynomials.  In particular, since any two polynomial models
on~$\mathcal F$ that differ by a confounding polynomial are
identical on $\mathcal F$, it is advantageous to choose models
from the quotient ring $R = k[x_1, \ldots , x_n]/I(\mathcal F)$
rather than from the polynomial ring itself.

It can be shown that the ring $R$ is isomorphic to the vector
space $k^s$, and we need to study possible vector space bases for
$R$ consisting of monomials. This can be done using Gr\"obner
bases of the ideal $I(\mathcal F)$ (see the appendix). For each
choice of a term order for $k[x_1,\ldots ,x_n]$, that is, a
special type of total ordering of all monomials, we obtain a
canonical generating set $G =\{g_1,\ldots ,g_s\}$ for $I(\mathcal
F)$. We obtain a canonical $k$-basis for the vector space $R\cong
k^s$ by choosing all monomials which are not divisible by the
leading monomial of any of the $g_i$.  We can then view each
polynomial in $R$ as a $k$-linear combination of the monomials in
the basis.

To be precise, let $\{T_1,\ldots , T_t\}$ be the set of all
monomials in the variables $x_1,\ldots ,x_n$ which are not
divisible by the leading monomial of any $g_i$.  Then each element
$f\in R$ can be expressed uniquely as a $k$-linear combination
$$
f = \sum_{j=1}^ta_jT_j,
$$
with $a_j\in k$. Suppose now that we are given a fractional design
$\mathcal F=\{\mathbf p_1,\ldots ,\mathbf p_r\}$ and an
experimental treatment resulting in values $f(\mathbf p_i)=b_i$
for $i=1,\ldots ,r$. If we now evaluate the generic polynomial~$f$
at the points~$\mathbf p_i$, we obtain a system of linear
equations
\begin{eqnarray*}
a_1T_1(\mathbf p_1)+\ldots +a_tT_t(\mathbf p_1)&=&b_1,\\
& \vdots & \\
a_1T_1(\mathbf p_r)+\ldots +a_tT_t(\mathbf p_r)&=&b_r.
\end{eqnarray*}
We can view these equations as a system of linear equations in the
variables~$a_j$ with the coefficients $T_j(\mathbf p_i)$. We now
obtain the main criterion for the unique identifiability of a
model $f$ from the fraction $\mathcal F$.

\begin{theorem}
\label{thm1} \cite[Thm. 4.12]{Ro} Let $\mathcal X =\{\mathbf
p_1,\ldots ,\mathbf p_r\}$ be a set of distinct points in $k^n$,
and let $f$ be a linear model with monomial support $\mathcal
S=\{T_1,\ldots ,T_t\}$, that is, $f=\sum_ia_iT_i$. Let $X(\mathcal
S,\mathcal X)$ be the $(r\times t)$-matrix whose $(i,j)$-entry is
$T_j(\mathbf p_i)$. Then the model $f$ is uniquely identifiable by
$\mathcal X$ if and only if $X(\mathcal S, \mathcal X)$ has full
rank.
\end{theorem}

In this section we have given a brief outline of a mathematical
framework within which one can use tools from computational
algebra to address the three experimental design problems listed
above.  In the next section we will describe a similar set of
problems and a similar approach to their solution in the context
of biochemical network modeling.
\section{Biochemical network inference}

Molecular biology has seen tremendous advances in recent years due
to technological breakthroughs that allow the generation of
unprecedented amounts and types of data.  For instance, it is now
possible to simultaneously measure the activity level of all genes
in a cell extract using DNA microarrays.  This capability makes it
possible to construct large-scale mathematical models of gene
regulatory and other types of cellular networks, and the
construction of such models is one of the central foci of
computational systems biology.  The availability of obtaining
experimental measurements for large numbers of entities that are
presumed to be interconnected in a network drives the need for the
development of network inference algorithms.  We will focus on the
mathematical aspects of this problem for the rest of the section.
More biological background can be found in \cite{LS}.

We consider a dynamic network with $n$ variables $x_1,\ldots ,x_n$.
These could represent products of $n$ genes in a cell extract from a
particular organism, say yeast. It is known that cellular metabolism
and other functions are regulated by the interaction of genes that
activate or suppress other genes and form a complex network. Suppose
we are given a collection of pairs of simultaneous measurements of
these variables:
$$
(\mathbf p_1,\mathbf q_1),\ldots , (\mathbf p_r,\mathbf q_r),
$$
with $\mathbf p_i,\mathbf q_i$ points in $\mathbf R^n$.  For gene
networks, each of these measurements could be obtained from a DNA
microarray.  Each pair $(\mathbf p_i,\mathbf q_i)$ is to be
interpreted as follows.  The variables in the network are
initialized at $\mathbf p_i$ and subsequently the network
transitions to $\mathbf q_i$.  This might be done through a
perturbation such as an experimental treatment, and $\mathbf p_i$
represents the network state immediately after the perturbation
and $\mathbf q_i$ represents the network state after the network
has responded to the perturbation. Sometimes the measurement pairs
are consecutive points in a measured time course.  In this case
the pairs above consist of consecutive time points.  Typically the
number $n$ of variables is orders of magnitude larger than the
number $r$ of measurements, in contrast to engineering
applications where the reverse is true (OR where $r$ is on the
order of $n$). For instance the network may contain hundreds or
thousands of genes, from which only 10 or 20 experimental
measurements are collected.

\begin{example}
\label{ex}
    Consider the following time course for a biochemical
    network of 3 genes, labeled $x_1, x_2,$ and $x_3$.
    \begin{center}
    \begin{tabular}{c|c|c}
      $x_1$ & $x_2$ & $x_3$ \\
      \hline
    1.91 & 3.30 & 1.98 \\
    1.50 & 1.42 & 1.99 \\
    1.42 & 1.31 & 0.03 \\
    0.83 & 1.96 & 1.01 \\
    0.97 & 2.08 & 1.01 \\
    \end{tabular}
    \end{center}
    Each gene's expression levels were measured at 5 consecutive time points
    and each entry represents a measurement. While the data are
    given in tabular form, we could have also represented the data as
    the pairs of network states
    \begin{eqnarray*}
      \left((1.91,  3.30,  1.98), (1.50,  1.42,  1.99)\right)\\
      \left((1.50,  1.42,  1.99),  (1.42,  1.31,  0.03)\right)\\
      \left((1.42,  1.31,  0.03),  (0.83,  1.96,  1.01)\right)\\
      \left((0.83,  1.96,  1.01),  (0.97,  2.08,  1.01)\right).
    \end{eqnarray*}
\end{example}

\medskip\noindent
{\bf Network inference problem.}  Given input-output measurements
$\{(\mathbf p_i,\mathbf q_i)\}$, infer a model of the network that
produced the data.

\medskip
One can consider a variety of different model types.  First it is
of interest to infer the directed graph of causal connections in
the network, possibly with signed edges indicating qualitative
features of the interactions.  Dynamic model types include systems
of differential equations, Boolean networks, Bayesian networks, or
statistical models, to name a few.  In light of the fact that DNA
microarray data contain significant amounts of noise and many
necessary parameters for models are unknown at this time, it
suggests itself to consider a finite number of possible states of
the variables $x_i$ rather than treating them as real-valued. This
is done by Bayesian network inference methods, for instance. The
issue of data discretization is a very subtle one.  On the one
hand, discrete data conform more to actual data usage by
experimentalists who tend to interpret, e.g., DNA microarray data
in terms of genes fold changes of regulation compared to control.
 On the other hand, a lot of information is lost in the process of
 discretizing data and the end result typically depends strongly
 on the method used.  In the extreme case, one obtains only two states
 corresponding to a binary ON/OFF view of gene regulation. In our case,
 a strong advantage of using discrete data is that it allows us to
 compute algorithmically the whole space of admissible models for a
 given data set, as described below.  Nonetheless, the result typically
 depends on the discretization method and much work remains to be done
 in understanding the effect of different discretization methods.
 Once the variables take on values in a finite set $k$ of states,
it is natural to consider discrete dynamical systems
$$
F:k^n\longrightarrow k^n.
$$
As mentioned, the dynamics is generated by repeated iteration of
the mapping $F$.  In order to have mathematical tools available
for model construction and analysis, one can make the assumption
that $k$ is actually a finite field rather than simply a set.  In
practice this is easily accomplished, since the only ingredient
required is the choice of a finite state set that has cardinality
a power of a prime number. With these additional assumptions our
models are \emph{polynomial dynamical systems}
$$
F =(f_1,\ldots ,f_n):k^n\longrightarrow k^n,
$$
with $f_\ell \in k[x_1,\ldots , x_n]$ for $\ell=1,\ldots ,n$.  (As
remarked above, any function from a finite set of points into a
field can be represented as a polynomial function.) The $\ell$-th
polynomial function $f_\ell$ describes the transition rule for
gene $x_\ell$ and hence $f_\ell$ is called the \emph{transition
function} for $x_\ell$.

Returning to the network inference problem, we can now rephrase it
in the following form: \emph{Given the state transitions $\{(\mathbf
p_i,\mathbf q_i)\}$, find a polynomial dynamical system (or polynomial model) $F$ such that $F(\mathbf
p_i)=\mathbf q_i$.}

\medskip
This problem can be solved one node at a time, that is, one
transition function at a time.  This ``local'' approach to inference
then begins with a collection $\{\mathbf p_i\}$ of points, and we
are looking for transition functions $f_\ell\in k[x_1,\ldots ,x_n]$
that satisfy the condition that $f_\ell(\mathbf p_i)=b_i$, where
$b_i$ is the $\ell$-th entry in $\mathbf q_i$.

\begin{example}
\label{ex-disc}
    Let
    \begin{eqnarray*}
      (\mathbf p_1, \mathbf q_1)&=&\left((2,2,2), (1,0,2)\right),\\
      (\mathbf p_2, \mathbf q_2)&=&\left((1,0,2),  (1,0,0)\right),\\
      (\mathbf p_3, \mathbf q_3)&=&\left((1,0,0),  (0,1,1)\right),\\
      (\mathbf p_4, \mathbf q_4)&=&\left((0,1,1),  (0,1,1)\right).
    \end{eqnarray*}
    be the discretization of the data in Example \ref{ex} into the 3-element field
    $k=\mathbb F_3$ by discretizing each coordinate separately, according to the
    method described in \cite{DVML}. Then the goal is to find a polynomial
    model $F:k^3\longrightarrow k^3$ such that
    $F(\mathbf p_i)=\mathbf q_i$ for $i=1,\ldots,4$. Since any such
    $F$ can be written as $F=(f_1,f_2,f_3)$,
    we can instead consider the problem of finding transition
    functions $f_\ell:k^3\longrightarrow k$ such that
    $f_\ell(\mathbf p_i)=\mathbf q_{i\ell},$ for all $1\leq \ell \leq 3$ and $1\leq i\leq 4$.
\end{example}

The similarity to the problem about experimental design in the
previous section is now obvious. Factors correspond to variables
$x_i$ representing genes; levels correspond to the elements of the
field $k$ representing gene states; the points $\mathbf p_i$ of
the factorial design correspond to experimental measurements; and
the $b_i$ in both cases are the same. As mentioned earlier, the
available experimental observations are typically much fewer than
the totality of possible system states. Thus, the objective in
both cases is the same: Find good polynomial models for the full
design from an experimental treatment of a fractional design.

The approach to a solution is quite similar as well.  Suppose we
are given two transition functions $f$ and $g$ that both agree on
the given experimental data, that is, $f(\mathbf
p_i)=b_i=g(\mathbf p_i)$ for all $i$. Then $(f-g)(\mathbf p_i)=0$,
so that any two transition functions differ by a polynomial
function that vanishes on all given observations, that is, by a
polynomial in the ideal of points $I(\mathbf p_1,\ldots ,\mathbf
p_r)$, which we called $I(\mathcal F)$ in the previous section.
If $f$ is a particular transition function that fits the data for
some $x_\ell$, then the space of all feasible models for $x_\ell$
is
$$
f+I(\mathbf p_1, \ldots ,\mathbf p_r).
$$

The problem then is to choose a model from this space. In design
of experiments, the single-variable monomials represent the
\emph{main effects} and the other monomials represent
\emph{interactions}.  In the biochemical network case the
situation is similar. Single-variable monomials in a model for a
gene regulatory network represent the regulation of one gene by
another, whereas the other monomials represent the synergistic
regulation of one gene by a collection of other genes, for example
through the formation of a protein complex. In general, very
little theoretical information is available about the absence or
presence of any given monomial in the model.  One possible choice
is to pick the normal form of $f$ with respect to a particular
Gr\"obner basis for the ideal $I(\mathbf p_1, \ldots ,\mathbf
p_r)$.  However, this normal form depends on the particular choice
of Gr\"obner basis.  Other approaches are explored in \cite{DJRS},
in particular an ``averaging" process over several different
choices of Gr\"obner basis.

\begin{example}
\label{ex-model}
    Returning to our running example, consider the following polynomials:
    \begin{eqnarray*}
      f_1(x_1,x_2,x_3) &=& 2x_2x_3+2x_2+2x_3, \\
      f_2(x_1,x_2,x_3) &=& 2x_3^3+x_2^2+x_2+2x_3+1, \\
      f_3(x_1,x_2,x_3) &=& 2x_3^2+2x_1+2.
    \end{eqnarray*}
    Each $f_\ell$ interpolates the
    discretized data for $x_\ell$ (see Example \ref{ex-disc}). The
    ideal of the input points $\mathbf p_1, \ldots ,\mathbf p_4$ is
    $$I=\langle x_1+x_2+2,  x_2x_3+2x_3^2+2x_1+x_2,
    x_2^2+2x_3^2+x_2+2x_3 \rangle.$$
    Then the model space for each $x_\ell$ is given by $f_\ell+I$.
    The Gr\"obner basis $G$ for $I$ with respect to the graded reverse lexicographical
    term order $\succ$ with $x_1\succ x_2\succ x_3$ is
    $$G=\{x_1+x_2+2,x_2x_3+2x_3^2+x_2+2x_3,x_2^2+2x_3^2+x_2+2x_3,x_3^3+2x_3\}.$$
    To choose a model for each $x_\ell$, we compute the normal form
    $\bar f_\ell$
    of $f_\ell$ with respect to $\succ$, resulting in the polynomial dynamical system
    $F=(\bar f_1,\bar f_2,\bar f_3):(\mathbb F_3)^3\longrightarrow (\mathbb F_3)^3$
    with
    \begin{eqnarray*}
      \bar f_1(x_1,x_2,x_3) &=& 2x_3^2 + x_3 \\
      \bar f_2(x_1,x_2,x_3) &=& x_3^2 +2x_3 + 1 \\
      \bar f_3(x_1,x_2,x_3) &=& 2x_3^2 + x_2 + 1.
    \end{eqnarray*}
\end{example}

Given a polynomial model $F=(f_1,\ldots ,f_n)$ for a network, one
can predict the connectivity structure of the nodes by analyzing the
relationship between the variables and the transition functions. For
example, the transition function for~$x_1$ given above is in terms
of~$x_3$, but not the other variables.  The interpretation is that
regulation of the gene represented by~$x_1$ is dependent only
on~$x_3$.  The dynamic behavior of the network can be simulated by
evaluating~$F$ on all possible network states, that is, on all
of~$k^n$.

\begin{definition}
    Let $F=(f_1,\ldots ,f_n):k^n\longrightarrow k^n$ be a polynomial
    dynamical system.  The \emph{wiring diagram} of $F$ is the directed graph
    $(V,E)$ with $V=\{x_1,\ldots,x_n\}$ and $E=\{(x_i,x_j):x_i \text{ is a variable of
    }f_j\}$. The \emph{state space} of $F$ is the directed graph
    $(V,E)$ with $V=k^n$ and $E=\{(\mathbf a,F(\mathbf a):\mathbf a\in k^n\}$.
\end{definition}

Viewing the structure and dynamics of a network via the wiring
diagram and state space, respectively, allows one to uncover
features of the network, including feedback loops and limit cycles,
respectively (for example, see \cite{LS}).

\begin{figure}[h]
    \includegraphics[width=0.9in,angle=90]{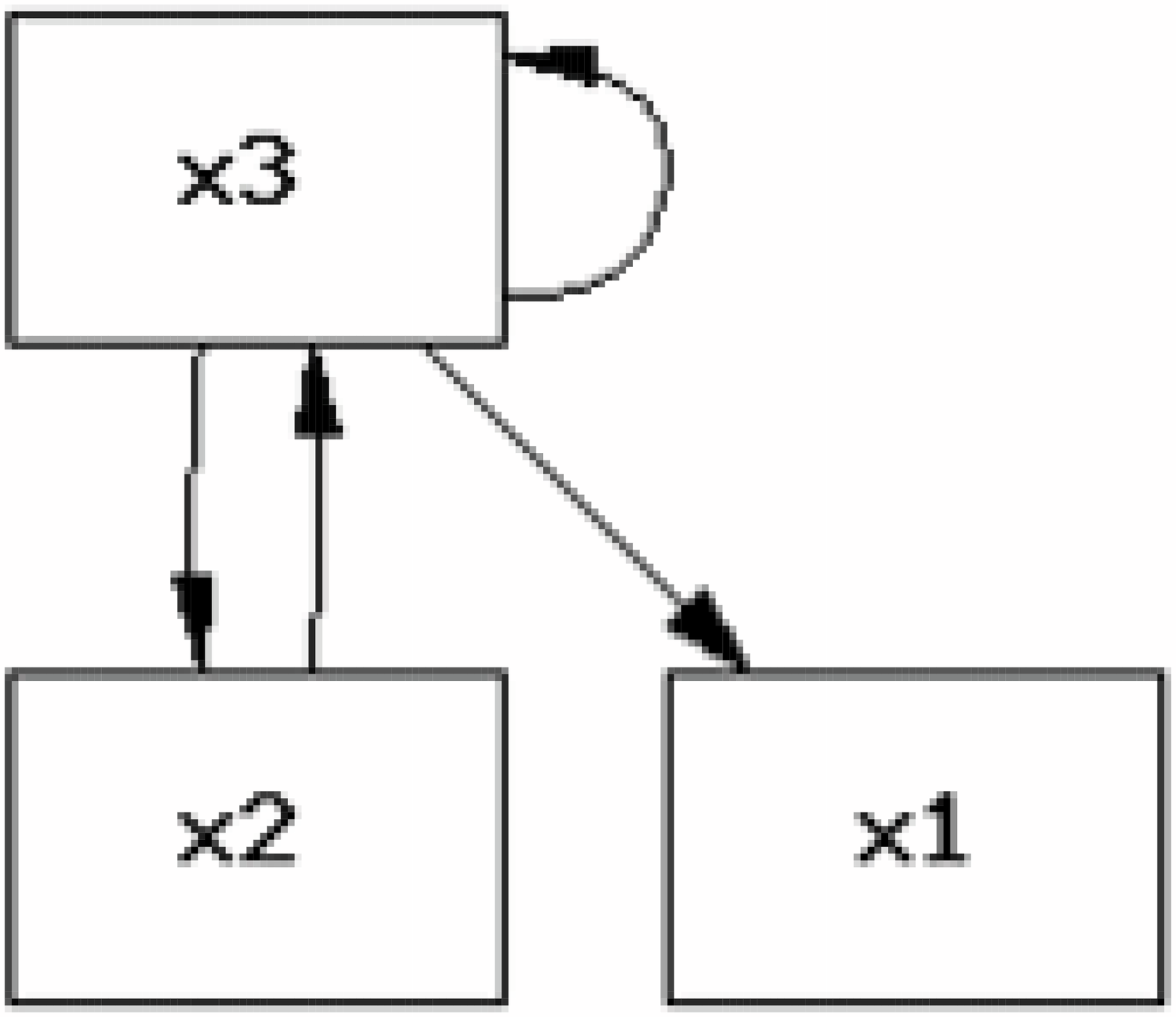}\\
    \includegraphics[width=1.5in,angle=90]{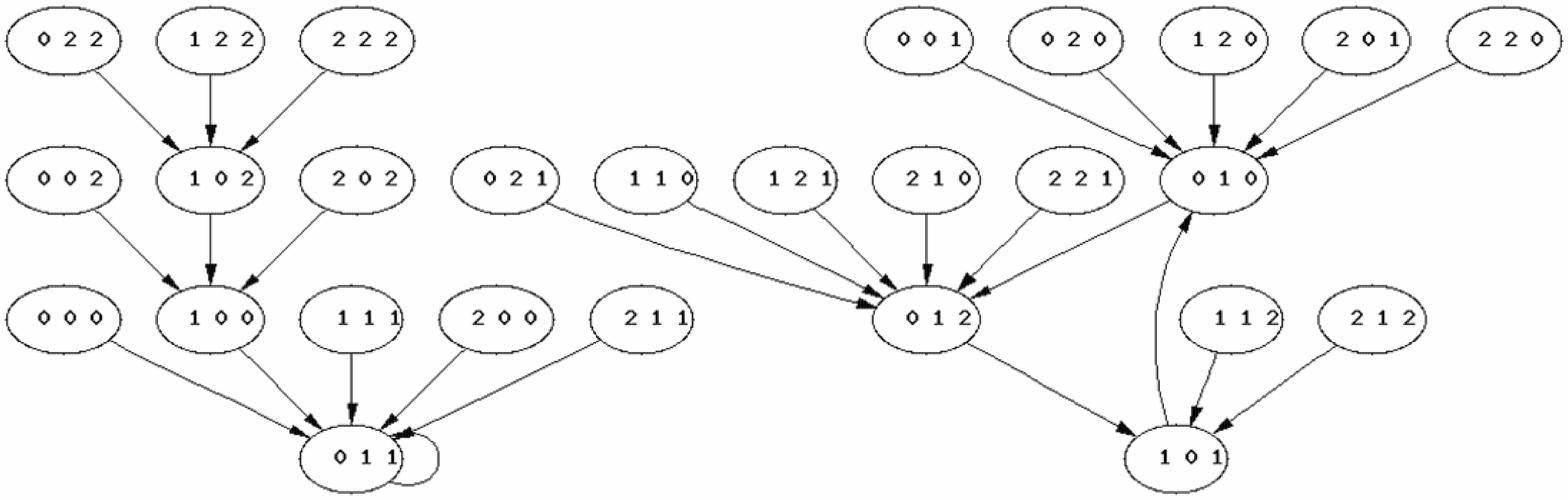}\\
    \caption{Wiring diagram (top) and state space (bottom)
    for the polynomial model $F$ in Example \ref{ex-model}.}
    \label{graphs}
\end{figure}

\begin{example}
    The polynomial model $F$ in Example \ref{ex-model} gives rise to
    the inferred wiring diagram and state space of the 3-gene network,
    as displayed in Figure \ref{graphs}. The network is predicted to
    have a feedback loop between $x_2$ and $x_3$, and the expression of $x_3$ is
    controlled via autoregulation.  Furthermore, the network has two possible
    limit cycles: the fixed point at (0,1,1) and the
    3-cycle on (0,1,0), (0,1,2), and (1,0,1).  The fixed point is
    considered to be an equilibrium state of the network, and
    the 3-cycle represents an oscillation.
\end{example}

While the above polynomial dynamical system may be a reasonable
model for the 3-gene network, it is not unique.  We recall from Theorem \ref{thm1}
that the number of monomials in the basis for
$k[x_1,x_2,x_3]/I(\mathbf p_1,\ldots ,\mathbf p_4)$
is the number of data points (4, in this case).  Since any transition function
can be written as a $k$-linear combination of the basis monomials,
then for a fixed term order there are $|k|^m=3^4$ possible
transition functions where $m$ is the number of data points.
In fact there are $(|k|^m)^n=3^{12}$ possible polynomial models,
given a term order. As there are 5 term orders which produce distinct
polynomial models
\footnote{We computed the marked Gr\"obner bases of the ideal
$I(\mathbf p_1,\ldots ,\mathbf p_4)$ via the Gr\"obner fan and
then computed the normal forms of the interpolating polynomials in
Example \ref{ex-model} with respect to each of these Gr\"obner bases
to obtain the 5 distinct polynomial models.}, there are $((|k|^m)^n)^5=3^{60}$
possible models for a 3-variable system on 3
states and 4 data points.

An important problem in this context that is common to both design
of experiments and biochemical network inference is the
construction of good fractional designs that narrow down the model
space as much as possible.  The challenge in network inference is
that experimental observations tend to be very costly, severely
limiting the number of points one can collect.  Furthermore, many
points are impossible to generate biologically or experimentally,
which provides an additional constraint on the choice of
fractional design.

\section{Polynomial dynamical systems}
It is worth mentioning that polynomial dynamical systems over
finite fields (not to be confused with dynamical systems given by
differential equations in polynomial form) have been studied in
several different contexts.  For instance, they have been used to
provide state space models for systems for the purpose of
developing controllers \cite{ML1, ML2} in a variety of contexts,
including biological systems \cite{JVDL}.  Another use for
polynomial dynamical systems is as a theoretical framework for
agent-based computer simulations \cite{LJMR}.  Note that this
class of models includes cellular automata and Boolean networks
(choosing the field with two elements as state set), so that
general polynomial systems are a natural generalization.  In this
context, an important additional feature is the update order of
the variables involved.

The dynamical systems in this paper have been updated in parallel,
in the following sense.  If $f=(f_1,\ldots ,f_n)$ is a polynomial
dynamical system and $\mathbf a\in k^n$ is a state, then
$f(\mathbf a)=(f_1(\mathbf a),\ldots ,f_n(\mathbf a))$. By abuse
of notation, we can consider each of the $f_i$ as a function on
$k^n$ which only changes the $i$th coordinate.  If we now specify
a total order of $1,\ldots , n$, represented as a permutation
$\sigma\in S_n$, then we can form the dynamical system
$$
f_\sigma = f_{\sigma(n)}\circ f_{\sigma(n-1)}\circ\cdots\circ
f_{\sigma(1)},
$$
which, in general, will be different from $f$.  Thus, $f_\sigma$
is obtained through \emph{sequential} update of the coordinate
functions. Sequential update of variables plays an important role
in computer science, e.g., in the context of distributed
computation.  See \cite{LJMR} for details.

Many processes that can be represented as dynamical systems are
intrinsically stochastic, and polynomial dynamical systems can be
adapted to account for this stochasticity. In the context of
biochemical network models, sequential update order arises
naturally through the stochastic nature of biochemical processes
within a cell that affects the order in which processes finish.
This feature can be incorporated into polynomial dynamical system
models through the use of random sequential update.  That is, at
each update step a sequential update order is chosen at random. It
was shown in \cite{cas05} in the context of Boolean networks that
such models reflect the biology more accurately than parallel
update models.  In \cite{Shmulevich:02b} a stochastic framework
for gene regulatory networks was proposed which introduces
stochasticity into Boolean networks by choosing at each update
step a random coordinate function for each variable, chosen from a
probability space of update functions.  Stochastic versions of
polynomial dynamical systems have yet to be studied in detail and
many interesting problems arise that combine probability theory,
combinatorics, and dynamical systems theory, providing a rich
source of cross-fertilization between these fields.
\section{Discussion}

This paper focuses on polynomial models in two fields, design of
experiments and inference of biochemical networks.  We have shown
that the problem of inferring a biochemical network from a
collection of experimental observations is a problem in the design
of experiments.  In particular, the question of an optimal
experimental design for the identification of a good model is of
considerable importance in the life sciences.  When focusing on
gene regulatory networks, it has been mentioned that conducting
experiments is still very costly, so that the size of a fractional
design is typically quite small compared to the number of factors
to be considered.  Another constraint on experimental design is
the fact that there are many limits to an experimental design
imposed by the biology, in particular the limited ways in which a
biological network can be perturbed in meaningful ways.  Much
research remains to be done in this direction.

An important technical issue we discussed is the dependence of
model choices on the term order used.  In particular, the term
order choice affects the wiring diagram of the model which
represents all the causal interaction among the model variables.
Since there is generally no natural way to choose a term order
this dependence cannot be avoided.  We have discussed available
modifications that do not depend on the term order, at the expense
of only producing a wiring diagram rather a dynamic model.  This
issue remains a focus of ongoing research.

As one example, an important way to collect network observations
is as a time course of measurements, typically at unevenly spaced
time intervals.  The network is perturbed in some way, reacts to
the perturbation, and then settles down into a steady state.  The
time scale involved could be on the scale of minutes or days.
Computational experiments suggest that, from the point of view of
network inference, it is more useful to collect several shorter
time courses for different perturbations than to collect one
highly resolved time course.  A theoretical justification for
these observations would aid in the design of time courses that
optimize information content of the data versus the number of data
points.

\section{Acknowledgements}
 Laubenbacher was partially supported by NSF Grant
 DMS-0511441 and NIH Grant R01 GM068947-01. Stigler was supported by
 the NSF under Agreement No. 0112050.

\bibliographystyle{siam}
\bibliography{Design_Exp_Laubenbacher_stigler}

\appendix

\section{Concepts from computational algebra}

In this section, we let $k$ denote a field and $R$ the polynomial
ring $k[x_1,\ldots,x_n]$. A subset $I\subset R$ is an \emph{ideal}
if it is closed under addition and under multiplication by elements
of~$R$.

\begin{definition}
Let $V$ be a finite set of points in $k^n$.  The set
$$\mathbf{I}(V)=\{f\in R:f(a)=0 \text{ for all } a\in V\}$$
of polynomials that vanish on $V$ is called the \textit{ideal of
points of~$V$}.
\end{definition}

Note that $\mathbf I(V)$ is indeed an ideal.  In fact, $\mathbf
I(V)$ is \emph{zero-dimensional} since the $k$-vector space
$R/\mathbf I(V)$ is finite dimensional with $dim_k(R/\mathbf
I(V))=|V|$.  While the number of generators of the vector space is
fixed, the generators themselves depend on the choice of \emph{term
order}.

\begin{definition}
    A \emph{term order} on
    $R$ is a relation $\succ$ on the set of
    monomials $\mathbf x^a:=x_1^{a_1}x_2^{a_2}\cdots x_n^{a_n}$
    such that $\succ$ is a total ordering, $$\mathbf x ^a\succ \mathbf x ^b
    \implies \mathbf x ^a\mathbf x ^c \succ \mathbf x ^b\mathbf x ^c$$
    for any monomial $\mathbf x^c$, and $\succ$ is a well-ordering;
    \textit{i.e.}, every nonempty subset
    of monomials has a smallest element under $\succ$.
\end{definition}

Given a term order $\succ$, every nonzero polynomial $f\in R$ has a
canonical representation as a formal sum of monomials
$$f=\sum_{i=1}^r a_if_i$$ with $f_i\in R$ and $a_i\in k$ for $i=1,\ldots, r$,
and $a_if_i\succ a_jf_j$ for all $i>j$. Moreover, $a_1f_1$ is called
the \emph{leading term} of~$f$.

\begin{definition}
    Let $\succ$ be a term order and $I\subset R$ an ideal.
    A finite subset $G=\{g_1,\ldots ,g_m\}\subset I$ is a
    \emph{Gr\"obner basis} for $I$ if the leading term of any
    $f\in I$ is divisible by the leading term of some $g_i$ under~$\succ$.
    The \emph{normal form} of~$f\in R$ with respect to~$G$, denoted $NF(f,G)$,
    is the remainder of~$f$ after division by the elements of~$G$.
\end{definition}


\begin{theorem}
    Every nonzero ideal $I\subset R$ has a Gr\"obner basis.
\end{theorem}

\begin{theorem}
    Let $G$ be a Gr\"obner basis for $I\subset R$ and let $f\in R$.
    Then $NF(f,G)$ is unique.
\end{theorem}

Let $G$ be a Gr\"obner basis and $LT(G)$ be the set of leading terms
of the elements of $G$. The set $\{\mathbf x^a: \mathbf x^a\notin
LT(G)\}$ is a basis for $R/\mathbf I(V)$ and its elements are called
\emph{standard monomials}. Given a Gr\"obner basis  $G$ of $\mathbf
I(V)$ with respect to a term order,  every nonzero polynomial $\bar
f\in R/\mathbf I(V)$ has a \emph{unique} representation as a formal
sum of the standard monomials.

\end{document}